\RequirePackage{fix-cm}
\documentclass[smallextended]{svjour3}       
\smartqed  
\usepackage{graphicx}
\begin{document}

\title{Investigation of weak-rates for odd-A nuclei  for presupernova simulations}


\author{Muhammad Riaz \and
       Jameel-Un Nabi \and
       Muhammad Majid
}


\institute{Muhammad Riaz\at
              Faculty of Sciences, Department of Physics, University of Okara, Okara,  Punjab, Pakistan.\\
              Tel.: +92-346-9462496\\
              \email{mriazgiki@gmail.com}           
           \and
           Jameel-Un Nabi\at
              University of Wah, Quaid Avenue, Wah Cantt 47040, Punjab, Pakistan\\
              GIK Institute of Engineering Sciences and Technology,\\ Topi 23640, Khyber
              Pakhtunkhwa, Pakistan.
}

\date{Received: date / Accepted: date}

\maketitle

\begin{abstract}
Calculating weak decay rates under stellar conditions for studying presupernova evolution of massive stars is a challenging task. Here we show the importance of odd-A nuclei for presupernova simulations. In order to calculate the required nuclear matrix elements we apply the pn-QRPA model in a deformed basis. Nuclear deformation, thought to play an integral role in calculation of associated weak decay rates, are taken into account in our model. We calculate Gamow Teller (GT) strength distributions, $\beta^-$ emission and positron capture rates for selected odd-A nuclei. Our model does not employ the Brink--Axel hypothesis as used in previous calculations of weak decay rates and we perform a state-by-state microscopic calculation of GT strength distributions from all parent excited states. Our calculated $\beta^-$ decay rates are in good agreement with large scale shell model calculations.
\keywords{Gamow-Teller transitions \and $\beta$ decay rates \and pn-QRPA theory \and presupernova}
\end{abstract}

\section{Introduction}
\label{intro}
The study of weak reaction rates are not only significant for particle and nuclear physics, but also are of much importance for nuclear astrophysics \cite{Ffn80}. The weak rates especially, lepton capture rates and $\beta$-decay processes are essential nuclear inputs for understanding the nucleosynthesis processes. In specific, the
Gamow-Teller (GT) transition rates, for highly excited levels of medium-heavy and heavy nuclide are significant constituent to model the late stellar evolution and the core-collapse supernova \cite{Lan03,Col12}. Alternatively, to constrain the neutrino mass and resolve the neutrino mass hierarchy in nuclear reactions such as neutrinoless double beta-decay, one requires to determine the nuclear matrix elements of GT strength between the
ground level of the parent and daughter nuclide as well as it is necessary to calculate
all the possible low- and high-lying states of the intermediate nuclide \cite{Avi08}. Majority of these weak rates properties cannot be obtained experimentally (although preliminary attempts have been
made under very exotic experimental conditions and, therefore, theoretical calculations are critical \cite{Fu15,Zha17}.\\
Majority of stellar nuclear reaction processes comprise of large number of unstable nuclei. Weak interaction properties of these nuclide play an important role in astrophysical processes. Substantial experimental work has been done in the last decades in order to determine the nuclear masses and many other properties of these exotic nuclide around the globe. Majority of these nuclide cannot be produced under laboratory conditions and theoretical estimates of weak-decay properties became more demanding to help us  understand these stellar processes. These charge-changing rates are the crucial constituents to be recognized in principally all stellar processes \cite{Bor06}. Stellar evolution and nucleosynthesis mechanism have been the attention of many calculations \cite{Arn96}. At the later phases of massive stars evolution, an iron core is ultimately formed and then no additional fuel exist for ignition of a new cycle. The astrophysical core progressively becomes unstable
and the core collapses because of capturing of leptons and photodisintegration processes. The core collapse is sensitive to lepton-to-baryon ratio ($Y_{e}$) of stellar matter \cite{Bet79}. The parameter $Y_{e}$ is governed by charge-changing reactions, i.e. lepton captures and $\beta$-decay. These charge-changing reactions significantly affect the late evolutionary stages of massive stars by altering the entropy of the core and $Y_{e}$  of the
presupernova star. The electron capture (EC) decreases the number of electrons available for pressure support, whereas the $\beta^-$decay play a key role thereby increasing the pressure support \cite{Hix03}.\\
Nuclear physicist are working hard to calculate, the ground as well as the excited-states GT strength distributions. Measurement of these strength is a challenging and difficult job, because huge number of nuclei are present inside the stellar matter. Due to the substantial implications of the charge-changing reaction rates in stellar scenario, they were extensively calculated using different nuclear models. The first significant effort to calculate the astrophysical rates were calculated by Fuller, Fowler and Newman (FFN) \cite{Ffn80}. They used the independent particle model and have employed the Brink–-Axel hypothesis \cite{Bri55} in their calculation.
The measured data present at that time was also incorporated in their calculation in order to enhance the reliability of their calculations. The FFN results were later expanded for
heavy nuclide ($A$ $>$ 60) by \cite{Auf94}.  Shell model Monte
Carlo (SMMC) technique \cite{Dea98} was employed for weak reaction rates calculation of \emph{fp}-shell nuclei. The benefit of this method is that it considers the nuclear temperature precisely, and also deal huge model spaces. On the other hand this  technique has issues in its applicability to odd–-odd and odd--$A$ nuclide at low temperatures. Furthermore, the shell model (SM) diagonalization may not be done beyond the fp-shell nuclide due to the larger dimensions of the involved model spaces. To solve this issue a hybrid model (SMMC + RPA) was presented \cite{Lan001,Sam03}. In this model, the nucleus is defined as a Slater determinant with temperature-dependent occupation numbers. In order to calculate the weak rates of hot nuclide, Dzhioev et al. \cite{Dzh09,Dzh10} employed thermal quasi-particle random-phase approximation (TQRPA) approach.\\
Cole et al. \cite{Col12} presented the capability of theoretical nuclear models to reproduce the measured GT strength of nuclear reactions at intermediate energies. The authors  concluded that the SM reproduce well the measured data, whereas QRPA calculations \cite{Mol90} showed larger deviations and overestimate the total experimental GT strength. It was also concluded that SM model calculated electron capture (EC) rates matches well the EC rates derived from the charge-exchange reactions as compared to those calculated by using the QRPA model.
Recently, the authors in Ref. \cite{Maj18} probed the conclusion of the Cole et al. study and suggested modifications in the
conclusions of Ref. \cite{Col12}. It was shown that 
QRPA calculations with a smart and optimum choice of model parameters may lead to data in very good comparison with measured values.

In the current manuscript we have calculated the $\beta$-decay rates and positron capture on odd-A nuclide ($^{45}$Sc and $^{55}$Mn) by applying the deformed pn-QRPA model. In the next
section, we briefly present the essential formalism of our model and its
parameters employed for the calculation of stellar weak rates.
Section 3 shows the result of our model calculations. Finally the conclusions are presented
in Section 4.

\section{Formalism}
The pn-QRPA formalism employed for computation of charge-changing
transitions and consequently for weak-interaction rates in this section.

The Hamiltonian considered in this model is specified by
\begin{equation} \label{GrindEQ__1_}
H^{pn-QRPA} =H^{sp} + V^{pairing} + V_{GT}^{ph} + V_{GT}^{pp} ,
\end{equation}
where $H^{sp}$ shows the single particle Hamiltonian, $V^{pairing}$
is pairing force (considered within the BCS approximation),
$V_{GT}^{ph}$ and $V_{GT}^{pp}$ represent the particle-hole
($ph$)and particle-particle ($pp$) interaction parameters for
Gamow-Teller strength, respectively. The single particle
eigenfunctions and eigenvalues were computed in Nilsson model
\cite{Nil55}, in which the nuclear deformation ($\beta_{2}$) is
taken. The $ph$ and $pp$ parameters are represented by $\chi$ and
$\kappa$, respectively. Our computed GT strength satisfy the model independent Ikeda
sum rule (ISR) \cite{Isr63}. Other parameters essential for
weak-interaction rates are the $\beta_{2}$, the pairing gap ($\Delta
_{nucleon}$), the Nilsson potential parameters (NPP) and the
Q-values. The NPP were chosen from \cite{Rag84} and the $\hbar
\omega = 41/A^{+1/3}$ in units of MeV is considered for Nilsson
oscillator constant, similar for baryons (N and P). The estimated
half-lives ($T_{1/2}$) values weakly rely on the pairing gaps
($\Delta _{nucleon}$) between nucleons \cite{Hir91}, are given as
\begin{equation}
\Delta _{n} =\Delta _{p} =12/\sqrt{A}  (MeV),
\end{equation}
$\beta_{2}$ was determined by using the formula
\begin{equation}
\beta_{2} = \frac {125 (Q_{2})} {1.44 (A)^{2/3} (Z)},
\end{equation}
where $Q_{2}$ denote the electric quadrupole moment chosen from
\cite{Mol16}. Q-values were chosen from the \cite{Aud12}.\\
In pn-QRPA model the charge-changing reaction transitions are
defined by phonon creation operators. The pn-QRPA phonons are given
as
\begin{equation}
A_{\omega}^{+}(\mu)=\sum_{pn}(X^{pn}_{\omega}(\mu)a_{p}^{+}a_{\bar{n}}^{+}-Y_{\omega}^{pn}(\mu)a_{n}
a_{\bar{p}}).
\end{equation}
Where the summation is taken on all the p-n pairs having $\mu$ =
\textit{m$_{p}$-m$_{n}$} = 0, $\pm$1, here
\textit{m$_{n}$}(\textit{m$_{p}$}) represent the angular momentum
third component of neutron(proton). The \textit{a$^{+}_{n(p)}$} show
the creation operator of a quasi-particle (q.p) state of
neutron(proton), however the \textit{$\bar{p}$} represents the time
reversed state of \textit{p}. The ground level of the theory with
respect to the QRPA phonons is considered as the vacuum,
A$_{\omega}(\mu)|QRPA\rangle$ = 0. The excitation energy ($\omega$)
and amplitudes (\textit{X$_{\omega}, Y_{\omega}$}) of phonon
operator are obtained by solving the famous RPA equation. Detailed
solution of RPA matrix equation can be seen in Refs. \cite{Hir91,Mut89}.\\
The decay of electron rates from the parent nuclide
\emph{nth} state transition to that of the daughter nuclei
\emph{mth} state is obtained as
\begin{equation}
\lambda _{nm}^{\beta-decay} =\ln 2\frac{f_{nm}(T,\rho
	,E_{f})}{(ft)_{nm}},
\end{equation}
the term $(ft)_{nm}$ is linked to the reduced transition probability
($B_{nm}$) by
\begin{equation}
(ft)_{nm} =D/B_{nm},
\end{equation}
The reduced transition probabilities $B_{nm}$'s are given by
\begin{equation}
B_{nm}=((g_{A}/g_{V})^{2} B(GT)_{nm}) + B(F)_{nm}.
\end{equation}
The $D$ value was chosen as 6143 s \cite{Har09} and $g_{A}/g_{V}$ was
considered as -1.254. The reduced Fermi and GT transition
probabilities are expressed as
\begin{equation}
B(F)_{nm} = \frac{1}{2J_{n} +1} \langle{m}\parallel\sum\limits_{k}
\tau_{\pm}^{k}\parallel {n}\rangle|^{2}
\end{equation}
\begin{equation}
B(GT)_{nm} = \frac{1}{2J_{n} +1} \langle{m}\parallel\sum\limits_{k}
\tau_{\pm}^{k}\overrightarrow{\sigma}^{k}\parallel {n}\rangle|^{2},
\end{equation}
where $\overrightarrow{\sigma}(k)$ and $\tau_{\pm }^{k}$ denote the
spin and the isospin operators, respectively. For the daughter and
parent excited level construction and computation of nuclear matrix
elements, within the deformed pn-QRPA theory, see \cite{Nab99}.

The phase space ($f_{nm}$) integral (over total energy) was
calculated (by adopting the natural units $\hbar=m_{e}=c=1$) as
\begin{equation}
f_{nm} = \int _{w_{1} }^{\infty }w\sqrt{w^{2} -1}(w_{m} -w)^{2} F(+
z, w) (1-R_{-})dw,
\end{equation}

In Eq.~(10) $w$ is the (kinetic energy + rest mass) of electron, however
$w_{l}$ is the electron threshold energy. $R_{-}$ shows the
distribution function of electron given as

\begin{equation}
R_{-} =\left[\exp \left(\frac{E-E_{f} }{kT} \right)+1\right]^{-1},
\end{equation}
where $E$ = ($w$ - 1),  $E_{f}$ and $k$  denote the kinetic energy and Fermi
energy of electrons and Boltzmann constant,
respectively. The Fermi functions denoted by ($F (+Z, w)$) were
estimated by the same way as in Ref. \cite{Gov71}. If the positron
(or electron) emission total energy ($w_{m}$) value is larger than
-1, then $w_{l}$ = 1, and if $w_{m}$ $\leq$ 1, then $w_{m}=|w_{l}|$,
$w_{m}$ is the total $\beta$-decay energy, and  is specified as
\begin{equation}
w_{m} = m_{p} -m_{d} + E_{n} -E_{m},
\end{equation}
\noindent where $E_{n}$ and $E_{m}$, represent parent and daughter excitation energies, respectively. As the
stellar core temperature is high enough so that there
is always a finite chance of occupation of parent excited levels.
The total electron decay and positron capture weak
rates were calculated as
\begin{equation}
\lambda^{\beta-decay} =\sum _{nm}P_{n} \lambda _{nm}^{\beta-decay}.
\end{equation}
In Eq.~13, the summation was applied on all final and initial states
until reasonable convergence in $\beta$-decay rates was obtained.

\section{Results and Discussions} \label{sec:results}
As per simulation results, the weak-interaction rates on medium heavy isotopes play a significant role during the late stages of core evolution. In our previous work Ref. \cite{Maj18}, we have calculated the charge-changing GT transitions strength for the odd-A (45Sc and 55Mn) medium-heavy nuclei in electron capture direction using the deformed pn-QRPA model. For stellar applications, the corresponding GT strength and capture rates were compared with previous theoretical and measured data.  It was concluded that our results were in decent comparison with the measured data. In this manuscript we have calculated the GT strength distribution, positron capture (PC) and $\beta^-$decay rates for odd-$A$ nuclei ($^{45}$Sc and $^{55}$Mn) by performing state-by-state GT calculations from all parent excited levels (considered 200 in this work) in pn-QRPA approach. Our model was able to take into account a large model space of 7$\hbar\omega$. Secondly, our nuclear model did
not utilize the Brink--Axel hypothesis \cite{BAH} (as employed in previous
computations) to evaluate GT transitions from parent
excited states. Brink’s hypothesis states that GT strength distribution on excited states is identical to that
	from ground state, shifted only by the excitation energy of the state. pn-QRPA approach, on the contrary, permits a microscopic state by
state estimation of GT charge-changing transitions from ground and
excited levels \cite{Nab20}.
Computed GT$^-$ strength of deformed pn-QRPA model for ground state of $^{45}$Sc and $^{55}$Mn  are shown in Fig.~1. The ordinates and abscissa of Fig.~1 shows the BGT$^{-}$ strength and daughter excitation energy up to 25 MeV respectively.  The left panel of Fig.~1 presents the computed BGT$^{-}$ strength for $^{45}$Sc nuclei, whereas the right panel shows the results for $^{55}$Mn nuclide. Fig.~1 depicts that the Gamow-Teller strength are well fragmented over a wide range of excitation energy for both cases of odd-$A$ nuclei. The centroid of $^{45}$Sc and $^{55}$Mn nuclide lies at 11.66 MeV and 17.82 MeV, respectively.

Fig.~2 represents the cumulative BGT$^{-}$ strength for ground state with respect to daughter states excitation energy and the total GT$^-$ strength for $^{45}$Sc and for $^{55}$Mn nuclei. The calculated total $\beta$-decay strength for $^{45}$Sc and $^{55}$Mn are 6.98 and 12.2, respectively.

Fig.~3 and Fig.~4 show the comparison of $\beta^-$decay rates in units of s$^{-1}$ in log$_{10}$ scale as a function of stellar temperature ($T_9$ denotes the temperature in units of 10$^9$ K) computed by FFN, LSSM and pn-QRPA models at three different densities $\rho Y_e=$10$^4$ gcm$^{-3}$, $10^8$ gcm$^{-3}$ and $10^{11}$ gcm$^{-3}$ (top, middle and bottom panels,  respectively) for $^{45}$Sc in Fig.~3 and for $^{55}$Mn in Fig. ~4. The figures show that all the rates calculated by three different models are, by and large, in good agreement with each other. However if one zooms-in then differences are seen which we feel could prove crucial for modeling and simulation of presupernova evolution of stars. We have computed $\beta^-$decay rates of pn-QRPA model by including all contribution from the excited states. These $\beta^-$decay rates for $^{45}$Sc and  $^{55}$Mn increase as the temperature increases, specially  in lower temperature region up to $T_9=$10 K then become saturated for both cases. As the density increases from $\rho Y_e=$10$^4$ gcm$^{-3}$ to $\rho Y_e=$10$^{11}$gcm$^{-3}$  $\beta^-$decay rates for $^{45}$Sc decrease. 

On comparison at lower temperature ($T_9$=0.7 K) and density ($\rho Y_e=$10$^4$ gcm$^{-3}$) calculated rates for $^{45}$Sc is -35.665 and at density ($\rho Y_e=$10$^8$ gcm$^{-3}$) is -72.315, respectively. For higher density at $\rho Y_e=$10$^{11}$gcm$^{-3}$ the rates were too low at $T_9=0.7$ K thats why we consider at $T_9=1.5$ K  is -89.978. In case of $^{55}$Mn rates are -60.419 at ($\rho Y_e=$10$^{4}$gcm$^{-3}$, $T_9=0.7$ K), -62.468 at ($\rho Y_e=$10$^{8}$gcm$^{-3}$, $T_9=0.2$ K) and -83.776 at ($\rho Y_e=$10$^{11}$gcm$^{-3}$, $T_9=1.5$ K), respectively. The above results suggest that as the mass number increases $\beta^-$decay rates increase.

Fig.~5 shows the calculated pn-QRPA total (sum of $\beta^-$decay and positron capture) weak interaction rates. These rates are shown as a function of stellar temperature at three different densities $\rho Y_e=$10$^4$ gcm$^{-3}$, $10^8$ gcm$^{-3}$ and $10^{11}$ gcm$^{-3}$ in each panel. The top panel shows total rates for $^{45}$Sc and bottom panel for $^{55}$Mn. For both cases as the density increases total weak mediated rates decrease, respectively. 

In case of  $^{45}$Sc at lower density ($\rho Y_e=$10$^4$ gcm$^{-3}$) total rates mediated at temperature ($T_9$=0.7 K) is -60.443, as density increases to $\rho Y_e=$10$^8$ gcm$^{-3}$ total rates mediated at ($T_9$=0.7 K) is -80.734 and by further increasing the density to  $\rho Y_e=$10$^{11}$ gcm$^{-3}$ the total rates become negligible at lower temperature, only -93.571 mediated rates at ($T_9$=3 K) can be counted. Similarly in the bottom panel of Fig.~5 mediated weak rates are enhanced in comparison to $^{45}$Sc because of higher mass number and shows the importance of heavy nuclei.

Table.~1 and Table.~2 show calculated positron capture (PC) and $\beta^-$decay rates for $^{45}$Sc and $^{55}$Mn, respectively at four different densities ($\rho Y_e=$10$^2$, 10$^5$, 10$^8$, 10$^{11}$ gcm$^{-3}$) and at selected temperatures in range of ($T_9=$0.1-30 K). The computed positron capture and $\beta^-$decay rates are highly sensitive to temperature. As the temperature increases the effective weak-decay rates get enhanced due to the fact that low-lying-states are populated.

\section{Conclusions} \label{sec:conclusions}
For presupernova evolution of massive stars  as well as simulation of type-II supernovae, $\beta^-$decay rates of $\emph{pf}$-shell nuclei play a crucial role. For assessment of these stellar $\beta^-$decay  rates it is important to compute charge changing Gamow-Teller transition strength distributions that may later be compared with measured data.

In this work we have calculated the GT strength distribution, PC and $\beta^-$decay rates of odd-A nuclei ($^{45}$Sc and $^{55}$Mn) by using deformed pn-QRPA model. Our model fulfilled well the model independent Ikeda sum rule. The stellar weak interaction rates are calculated over wide range of stellar densities (10 -- 10$^{11}$ g/cm$^{3}$) and temperature ($T_9$ = 0.01 -- 30). It was noted that the PC and $\beta^-$decay rates are highly sensitive to temperature and increased with core temperature. We compared our results with the previous work of FFN and LSSM. Our results are in overall reasonable agreement with previous theoretical calculations. However finite differences are seen, albeit of small magnitude. We feel that these small differences might fine-tune the modeling and simulation results and core-collapse supernova simulators are urged to test-run our calculated weak decay rates for possible interesting outcomes. 

\begin{acknowledgements}
J.-U. Nabi would like to
acknowledge the support of the Higher Education Commission
Pakistan through project numbers  5557/KPK/NRPU/R$\&$D/HEC/2016 and  9-5(Ph-1-MG-7)Pak-Turk/R$\&$D/HEC/2017 and
Pakistan Science Foundation through project number
PSF-TUBITAK/KP-GIKI (02).
\end{acknowledgements}

\begin{figure}
	\includegraphics [width=1\textwidth]{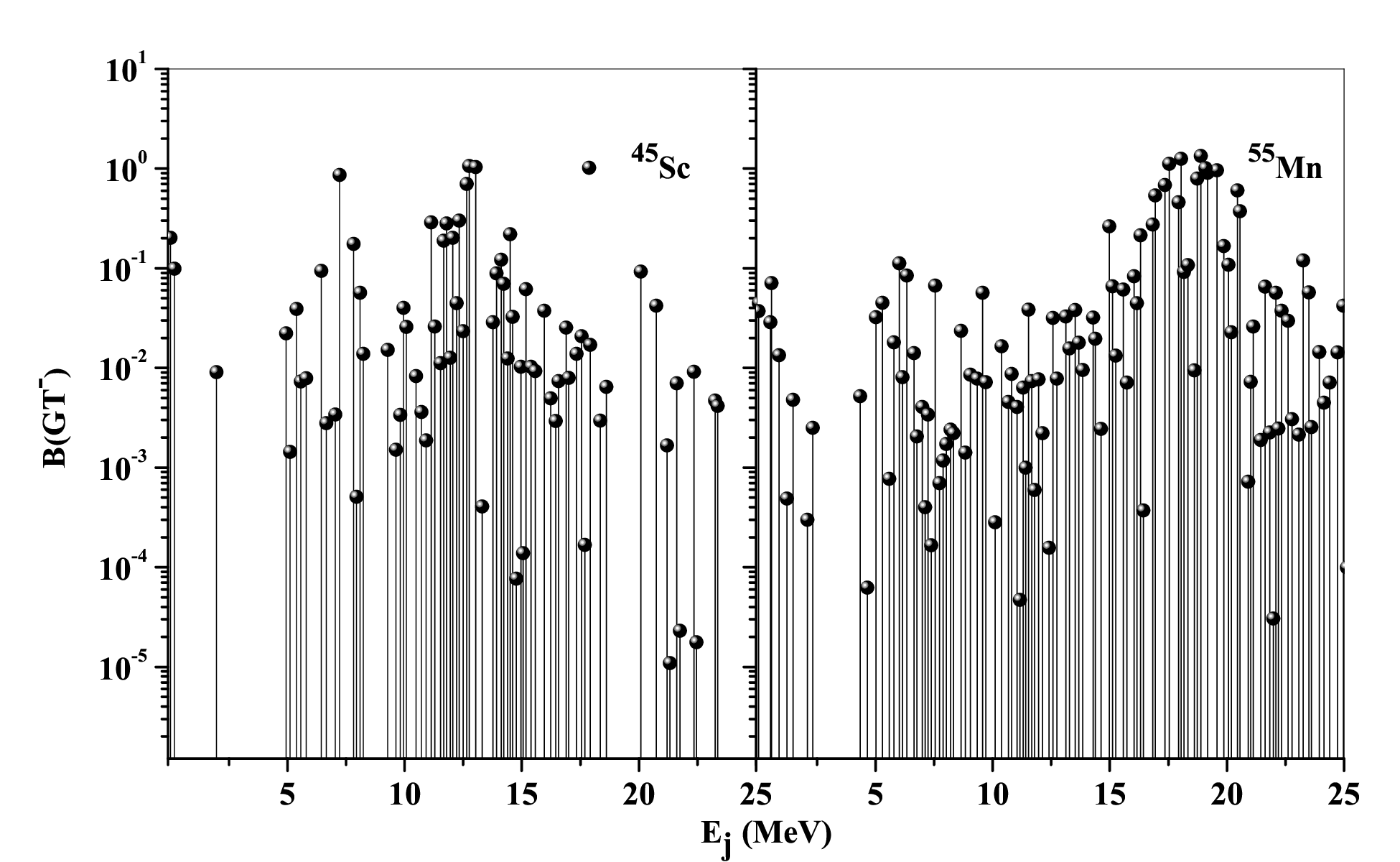}
	\centering \caption{ GT transitions from $^{45}$Sc to $^{45}$Ti (left panel) and $^{55}$Mn to $^{55}$Fe (right Panel)
		plotted as a function of the excitation energy of the daughter nucleus. }\label{fig1}
\end{figure}

\begin{figure}
	\includegraphics [width=1\textwidth]{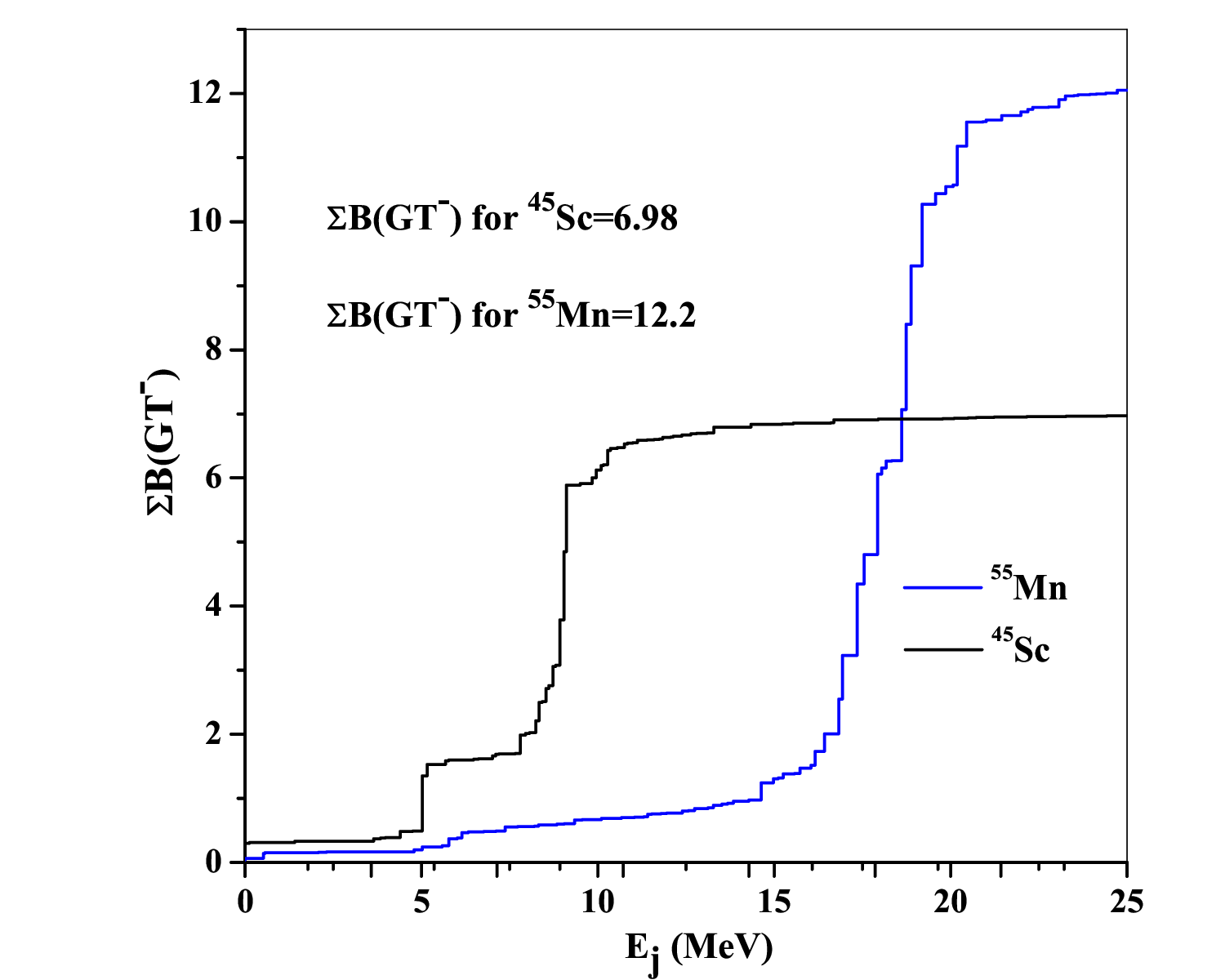}
	\centering \caption{ Comulative GT transitions strength from $^{45}$Sc to $^{45}$Ti and $^{55}$Mn to $^{55}$Fe. The abscissa shows  the excitation energy of the daughter nucleus. }\label{fig2}
\end{figure}

\begin{figure}
	\includegraphics [width=1\textwidth]{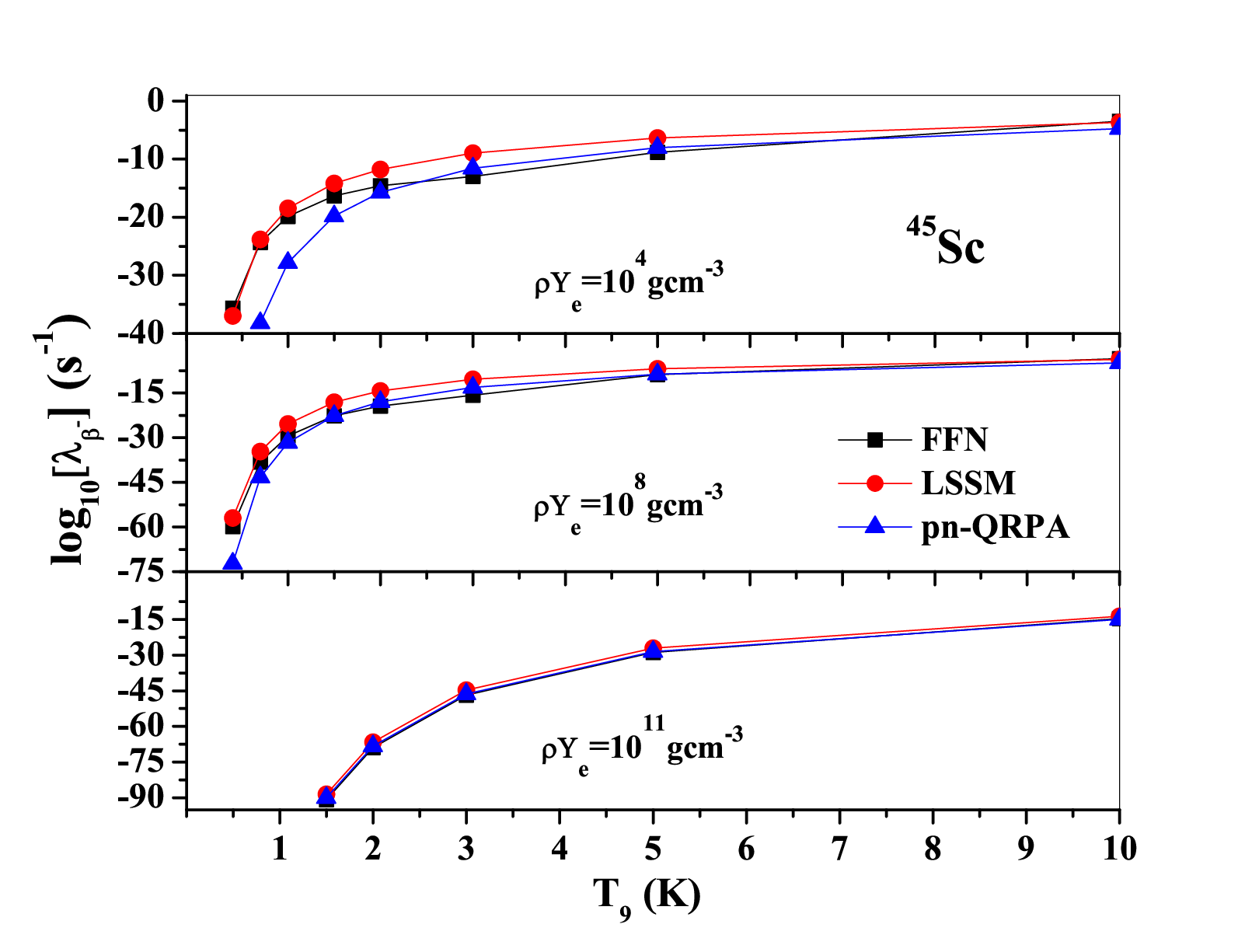}
	\centering \caption{Comparison of pn-QRPA calculated $\beta^-$decay rates for $^{45}$Sc with FFN and LSSM at three different selected densities as a function of core temperature. Stellar rates and temperature are in units of $s^{-1}$ and 10$^9$K, respectively.
	}\label{fig3}
\end{figure}

\begin{figure}
	\includegraphics [width=1\textwidth]{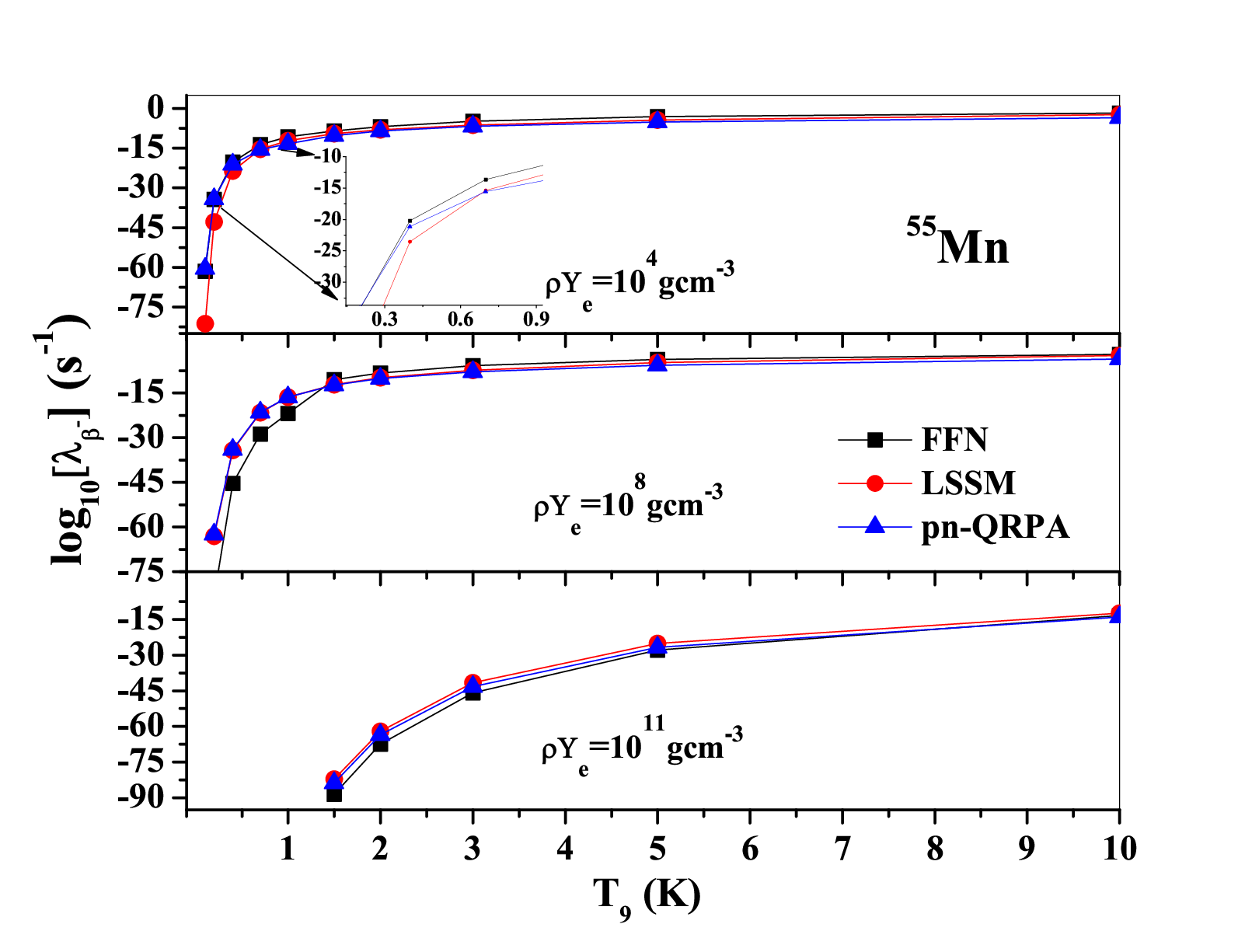}
	\centering \caption{Same as Fig~3 but for $^{55}$Mn.}\label{fig4}
\end{figure}

\begin{figure}
	\includegraphics [width=1\textwidth]{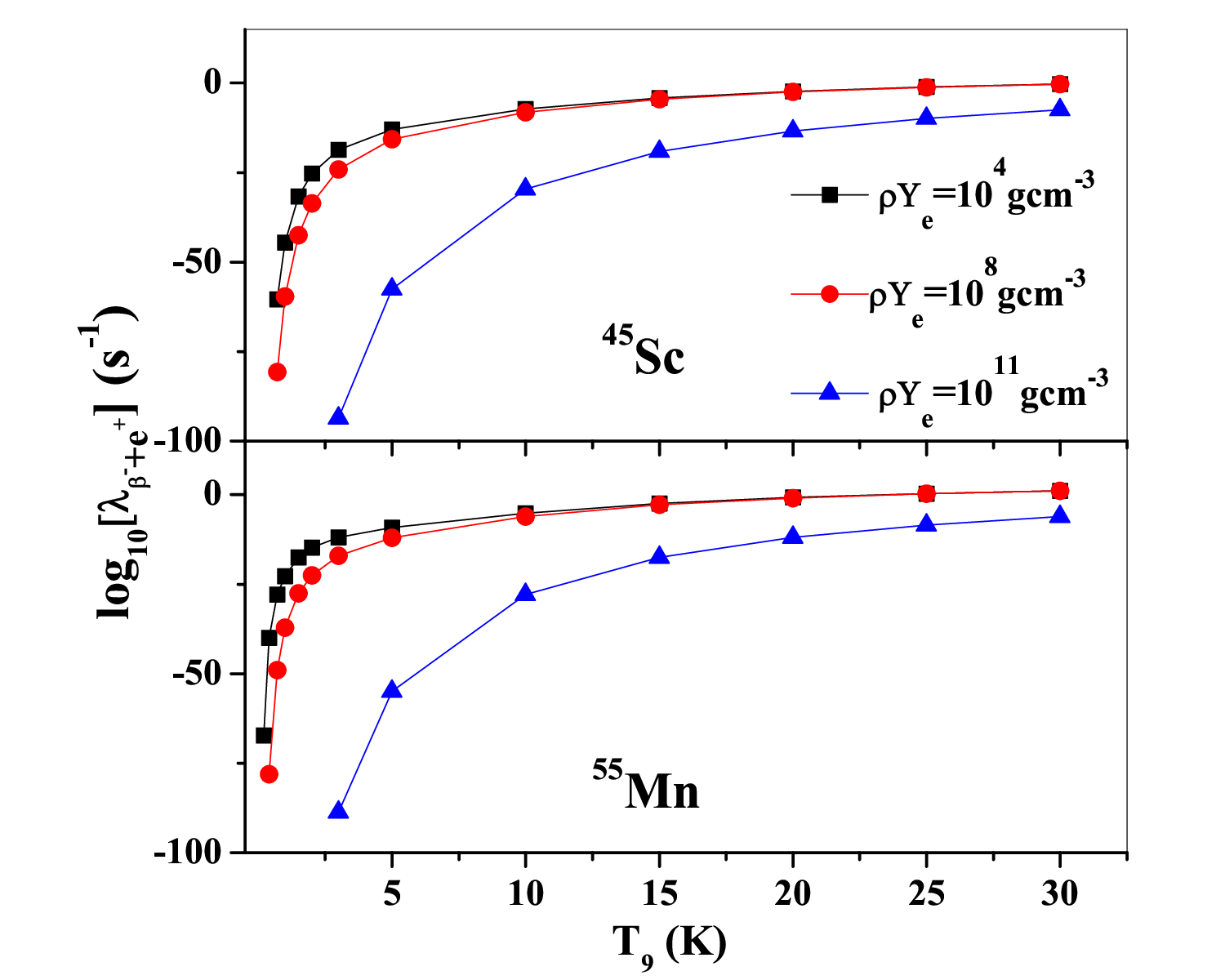}
	\centering \caption{Comparison of total rates,  $\lambda_{\beta^-+e^+}$, at three different densities as a function of core temperature for $^{45}$Sc (top panel) and for $^{55}$Mn (bottom panel).}\label{fig5}
\end{figure}

\clearpage
%
\begin{table}
	\scriptsize \caption{Calculated positron capture (PC) and electron
		decay (E-Decay) rates in stellar matter for $^{45}$Sc as a function of
		stellar density and temperature.  $\rho$Y$_{e}$ shows the stellar
		density (in units of g/cm$^{3}$) and T$_{9}$ gives the temperature
		in units of 10$^{9}$ K. The calculated rates are tabulated in log to
		base 10 scale and given in units of s$^{-1}$.} \label{Table 1}
	\begin{center}
		\begin{tabular} {cc|cc|cc|cc}
			\hline\hline
			$\rho$$\it Y_{e}$ & T$_{9}$ & PC    & $\beta$-Decay & $\rho$$\it Y_{e}$ & T$_{9}$ &  PC    & $\beta$-Decay \\
			\hline
			
			10$^{2}$ & 0.7   & -20.322 & -38.163 & 10$^{8}$ & 0.7   & -37.433 & -43.301 \\
			10$^{2}$ & 1     & -15.698 & -27.836 & 10$^{8}$ & 1     & -27.954 & -31.692 \\
			10$^{2}$ & 1.5   & -11.770 & -19.789 & 10$^{8}$ & 1.5   & -19.912 & -22.625 \\
			10$^{2}$ & 2     & -09.501 & -15.725 & 10$^{8}$ & 2     & -15.564 & -18.005 \\
			10$^{2}$ & 3     & -07.064 & -11.559 & 10$^{8}$ & 3     & -11.019 & -13.106 \\
			10$^{2}$ & 5     & -04.880 & -08.023 & 10$^{8}$ & 5     & -07.089 & -08.657 \\
			10$^{2}$ & 10    & -02.460 & -04.776 & 10$^{8}$ & 10    & -03.210 & -04.922 \\
			10$^{2}$ & 15    & -00.675 & -03.525 & 10$^{8}$ & 15    & -00.948 & -03.587 \\
			10$^{2}$ & 20    &  00.556 & -02.910 & 10$^{8}$ & 20    &  00.439 & -02.942 \\
			10$^{2}$ & 25    &  01.424 & -02.563 & 10$^{8}$ & 25    &  01.364 & -02.581 \\
			10$^{2}$ & 30    &  02.075 & -02.347 & 10$^{8}$ & 30    &  02.041 & -02.358 \\
			
			\hline
			10$^{5}$ & 0.7   & -23.343 & -38.193 & 10$^{11}$ & 0.7   & -100.00 & -100.00 \\
			10$^{5}$ & 1     & -17.758 & -27.860 & 10$^{11}$ & 1     & -100.00 & -100.00 \\
			10$^{5}$ & 1.5   & -12.660 & -19.804 & 10$^{11}$ & 1.5   & -92.178 & -89.978 \\
			10$^{5}$ & 2     & -09.825 & -15.732 & 10$^{11}$ & 2     & -69.804 & -68.210 \\
			10$^{5}$ & 3     & -07.123 & -11.562 & 10$^{11}$ & 3     & -47.257 & -46.314 \\
			10$^{5}$ & 5     & -04.890 & -08.024 & 10$^{11}$ & 5     & -28.979 & -28.619 \\
			10$^{5}$ & 10    & -02.461 & -04.777 & 10$^{11}$ & 10    & -14.468 & -15.109 \\
			10$^{5}$ & 15    & -00.675 & -03.525 & 10$^{11}$ & 15    & -08.636 & -10.470 \\
			10$^{5}$ & 20    &  00.556 & -02.910 & 10$^{11}$ & 20    & -05.368 & -08.092 \\
			10$^{5}$ & 25    &  01.424 & -02.563 & 10$^{11}$ & 25    & -03.268 & -06.647 \\
			10$^{5}$ & 30    &  02.076 & -02.347 & 10$^{11}$ & 30    & -01.786 & -05.676 \\
			\hline\hline
		\end{tabular}
	\end{center}
\end{table}
\begin{table}
	\scriptsize \caption{Same as Table~2, but for $^{55}$Mn.}
	\label{Table 2}
	\begin{center}
		\begin{tabular} {cc|cc|cc|cc}
			\hline\hline
			$\rho$$\it Y_{e}$ & T$_{9}$ & PC & $\beta$-Decay & $\rho$$\it Y_{e}$ & T$_{9}$ &  PC & $\beta$-Decay \\
			\hline
			
			10$^{2}$ & 0.7   & -10.391 & -15.565 & 10$^{8}$ & 0.7   & -27.502 & -21.555 \\
			10$^{2}$ & 1     & -08.447 & -13.280 & 10$^{8}$ & 1     & -20.704 & -16.444 \\
			10$^{2}$ & 1.5   & -07.084 & -10.329 & 10$^{8}$ & 1.5   & -15.225 & -12.338 \\
			10$^{2}$ & 2     & -06.276 & -08.507 & 10$^{8}$ & 2     & -12.336 & -10.197 \\
			10$^{2}$ & 3     & -05.252 & -06.678 & 10$^{8}$ & 3     & -09.203 & -07.906 \\
			10$^{2}$ & 5     & -04.040 & -05.140 & 10$^{8}$ & 5     & -06.241 & -05.752 \\
			10$^{2}$ & 10    & -01.723 & -03.447 & 10$^{8}$ & 10    & -02.467 & -03.620 \\
			10$^{2}$ & 15    &  00.112 & -02.545 & 10$^{8}$ & 15    & -00.156 & -02.612 \\
			10$^{2}$ & 20    &  01.269 & -02.044 & 10$^{8}$ & 20    &  01.153 & -02.076 \\
			10$^{2}$ & 25    &  02.064 & -01.752 & 10$^{8}$ & 25    &  02.005 & -01.770 \\
			10$^{2}$ & 30    &  02.657 & -01.570 & 10$^{8}$ & 30    &  02.623 & -01.582 \\
			
			\hline
			
			10$^{5}$ & 0.7   & -13.412 & -15.631 & 10$^{11}$ & 0.7   & -100.00 & -100.00 \\
			10$^{5}$ & 1     & -10.507 & -13.319 & 10$^{11}$ & 1     & -100.00 & -100.00 \\
			10$^{5}$ & 1.5   & -07.973 & -10.333 & 10$^{11}$ & 1.5   & -87.491 & -83.776 \\
			10$^{5}$ & 2     & -06.598 & -08.509 & 10$^{11}$ & 2     & -66.576 & -63.577 \\
			10$^{5}$ & 3     & -05.311 & -06.679 & 10$^{11}$ & 3     & -45.440 & -43.243 \\
			10$^{5}$ & 5     & -04.049 & -05.140 & 10$^{11}$ & 5     & -28.131 & -26.792 \\
			10$^{5}$ & 10    & -01.724 & -03.447 & 10$^{11}$ & 10    & -13.723 & -14.137 \\
			10$^{5}$ & 15    &  00.112 & -02.545 & 10$^{11}$ & 15    & -07.840 & -09.637 \\
			10$^{5}$ & 20    &  01.269 & -02.044 & 10$^{11}$ & 20    & -04.647 & -07.292 \\
			10$^{5}$ & 25    &  02.064 & -01.752 & 10$^{11}$ & 25    & -02.620 & -05.867 \\
			10$^{5}$ & 30    &  02.657 & -01.570 & 10$^{11}$ & 30    & -01.197 & -04.915 \\	
			\hline\hline
		\end{tabular}
	\end{center}
\end{table}

\end{document}